%% definitions for symbols, etc.
\def\gtapp
{\mathrel{\hbox{\raise0.3ex\hbox{$>$}\kern-0.8em\lower0.8ex\hbox{$\sim$}}}}
\def\ltapp
{\mathrel{\hbox{\raise0.3ex\hbox{$<$}\kern-0.75em\lower0.8ex\hbox{$\sim$}}}}

%%%%%%%%%

\magnification=1200

\centerline{\bf GAMMA-RAY BURSTS AS A COSMIC WINDOW FOR GALAXY EVOLUTION}

\vskip .2in

\centerline {I.F. Mirabel$^{1,2}$, D.B. Sanders$^3$,  E. Le Floc'h$^{1,4}$}

\vskip .1in

\noindent $^1$CEA/DSM/DAPNIA/SAp. Centre d'Etudes de Saclay. 91191. Gif/Yvette. 
France
\vskip .05in

\noindent$^2$Intituto de Astronom\'\i a y F\'\i sica del Espacio. Conicet. Argentina

\vskip .05in

\noindent$^3$Institute for Astronomy, University of Hawaii, 2680 Woodlawn Dr., 
Honolulu, HI 96822

\vskip .05in

\noindent$^4$European Southern Observatory. Casilla 19001, Santiago 19. Chile

\vskip .2in
{\bf Abstract: Present knowledge indicates that gamma-ray bursts 
are linked with massive stars. They will become invaluable probes of the 
early universe and galaxy formation. In the future, it will be possible 
to use gamma-ray bursts for two purposes: 
1) to probe the history of massive star formation in the Universe by 
the rate of occurence of 
gamma-ray bursts, and 2) for the study of galaxy evolution 
at all lookback times by determining the nature of the galaxy hosts. 
Because gamma-rays are not attenuated by intervening dust and gas, 
the selection of the cosmic sites of 
massive star formation by this method is less affected by
the biases associated with optical-uv surveys (e.g. UV-dropout 
techniques). Infrared and sub-millimeter follow up studies of the hosts of 
gamma-ray bursts may: 1) reveal a putative population of reddened 
($R-K \geq 4$) galaxies at high redshifts, and 2) detect very massive 
stars (population III) formed at $z \geq$ 5.}

\vskip .1in

\centerline{\bf Dusty starbursts and AGNs at high redshift}

\vskip .05in

Optical surveys with the Hubble Space Telescope and large ground based telescopes 
have enabled in recent years 
the study of luminous UV galaxies out to 
redshifts z$\sim$4 (Madau et al. 1996; Steidel et al. 1998). From the  
early analysis of galaxies in the HDF, Madau et al. (1996) concluded that 
there seemed to be little evidence for substantial amounts of dust obscuration 
in high redshift galaxies. It is remarkable how quickly this picture has 
changed  after the recent deep field surveys in the mid-infrared, far-infrared, 
and submillimiter wavelengths (e.g. Smail et al. 1997; Hughes et al. 1998; Barger et al. 1998; Blain et al. 1999; Elbaz et al. 1999). 
These new surveys have discovered 
a strongly evolving population of Luminous Infrared Galaxies (see Sanders \& Mirabel 1996, for a review) 
that could be the progenitors of the present-day population of massive spheroidal galaxies.

The importance of dust at high redshifts has also been
pointed out by Fabian (1999), who proposes a large population
of dust enshrouded AGNs at $z>1$ to account for the X-ray background.
This prediction has recently been confirmed by the discovery with Chandra of 
a large population of X-ray point sources with ``extremely faint, or in some cases 
undetectable" (i.e. $I > 26$) `` optical counterparts" (Mushotzky 
et al. 2000). Furthermore, at millimeter and submillimeter wavelengths thermal 
emission from dust is currently detected from quasars up to redshifts of 5 
(e.g. Omont et al. 1996; Carrili et al. 2000). The implied far-infrared 
luminosities are $\gtapp 10^{13} L_\odot$, dust masses 
$\gtapp 10^8 M_\odot$, and molecular gas masses of a few $\times 10^{10} M_\odot$. 
These major events in galaxy evolution at high redshifts are largely missed 
by UV/optical surveys, and it is now clear that the use of such 
surveys alone may lead to a subtantial underestimate of the global star 
formation rate and certainly to a distorted picture of the early evolution 
of the most massive galaxies. What may be the fraction of massive star formation 
in the universe that occured in galaxies so heavily obscured by dust that could 
not be detected in UV-selected surveys is an issue of current debate (e.g 
Adelberger \& Steidel 2000; Sanders 2000).

\vskip .05in

\centerline{\bf Gamma-ray bursts: A new cosmic window for galaxy evolution} 

\vskip .05in

Another major timely event in astrophysics has been the dramatic 
tranformation in our 
understanding of gamma-ray bursts (GRBs). There is increasing 
evidence that the most common GRBs, those with durations longer than 
a few seconds, are at cosmological distances and 
are associated with sites of massive star formation. 
The physical properties of the afterglows, their locations at a few kpc from 
the center of 
host galaxies (e.g. Djorgovski 1999), and the 
statistics from the several thousands of GRBs detected so far 
with BATSE (Fishman et al. 1999), give strong 
support to the idea that the majority of GRBs are linked to the cataclysmic collapse 
of massive stars into black holes (Paczy\'nski 1998; MacFayden \& Woosley 1999). 
In this context, GRBs can be used as sign-posts to probe more accurately the history 
of massive star formation and galaxy evolution. 

One of the great advantages of this approach is that GRBs can be detected 
at all lookback times. The 
broad band spectral and temporal behavior of the afterglows have confirmed 
that the bursts are beamed in relativistic jets, with the observer near the 
jet axis (M\'esz\'aros 1999; Kulkarni et al. 1999; Galama et al. 1999; 
Castro-Tirado et al. 1999). This leads to observed energies that are 
Doppler boosted by several orders of magnitude (Mirabel \& Rodr\'\i guez 
1999). Therefore, the detection of GRBs depends essentially on the 
beaming 
angle rather than on their distance, and they can be observed up to very large 
redshifts. An example of this was the optical afterglow of GRB990123, which  
could 
be seen with a simple pair of binoculars, since it reached an optical brightness 
of 8.6 mag, despite the fact that it came from z=1.6. High resolution 
spectroscopy of such optically bright bursts, which are many magnitudes brighter 
than quasars at the same redshift, could be very valuable to probe with 
unprecedented sensitivity the Ly$\alpha$ forest, and consequently, the chemical 
evolution of the universe.

The nine GRBs with redshifts determined by {\it optical} techniques 
prior to the end of 1999 are in the redshift range $z = 0.43-3.47$, 
with a mean of $z=1.3$ 
(see also Blain \& Natarajan 1999). However, this mean redshift is almost 
certainly biased toward low values because sources at 
very early epochs appear rather red and it is difficult 
to determine their redshifts by optical observations alone. For instance, 
no [OIII] 
or Ly${\alpha}$ lines from objects at $1.3 \leq z \leq 2.5$ can be seen 
in the optical.  Furthermore, there are several GRB afterglows that appeared 
to be extremely red whose redshifts are unknown. Some of these could be related 
to the dead of a putative class of Population III stars, which are very massive 
stars brought on by cooling of molecular hydrogen in the dark ages of 
the very high 
redshift universe, before galaxies were formed (Lamb \& Reichart 1999; 
Bromm, Coppi \& Larson 1999). 
One of the best cases to explore this possibility is GRB980329, for which 
Fruchter (1999) hypothesized $z \geq 5$.

Dust obscuration is very important in studies of recently formed compact 
objects (e.g. Mirabel et al. 1999). Because GRBs are the last phase 
of the evolution of the most massive stars (Woosley 1999) 
which do not live long enough to leave their
place of birth, it is expected that a 
large fraction will be still enshrouded in their placental clouds of 
molecular gas and dust. The observations of GRB  
afterglows are consistent with this picture: 1) among the 47 X-ray afterglows 
with good localizations, {\it about half} had no optical counterparts 
(Greiner 1999),  2) despite persistent optical follow ups, three radio 
afterglows without optical counterparts have been detected, and 
3) some afterglows  
appear to be extremely reddened (e.g. GRB990705). 
Dust obscuration is the most likely reason for the non detection of 
optical transients. Indeed the 
absence of detectable optical flux accompanying strong X-ray emission in the 
1-60 keV energy band of 
SAX may simply be due to the strong redshift dependence 
of dust obscuration at optical wavelengths in comparison to the X-rays, 
i.e. $\tau_{\rm opt}$/$\tau_{\rm X-ray}$ $\propto$ $(1+z)^4$ (Taylor et al. 1998). 
Therefore, GRBs can be used to detect a putative population of dust-enshrouded 
forming galaxies at high redshifts.

\vskip .05in

\centerline{\bf The future}

\vskip .05in   
After the breakthrough produced by the X-ray satellite Beppo-SAX 
(Costa et al. 1997), it is 
expected that in the comming years, high energy space missions 
(SAX, HETE2, CHANDRA, XMM, INTEGRAL and SWIFT) 
will be providing about 50 GRB X-ray afterglows per year 
with localizations in the range of 1 arcsec to 1 arcmin. Furthermore, 
the localizations of the 
gamma-ray bursts can be improved a posteriori using the Interplanetary 
Network. 

We note 
that the selection of the initial sample of star-formation sites in this approach 
is less affected 
by the problems and biases associated with the optical-UV selections of star 
formation sites, which are strongly influenced by reddening effects. Indeed 
the gamma rays are not attenuated by the 
intervening columns of gas and dust.  The bursts thus provide an unbiased 
sampling of cosmic sites of massive star-formation. 

An important caveat is that most X-ray 
afterglows until now have had error boxes of a few arcmin$^2$, a region large 
enough to contain from
tens up to a few hundred galaxies, therefore it is very difficult without 
afterglow detections in the optical, 
infrared or radio wavelengths to identify unambigously which object is 
the host of the GRB. However, the science programs of the recently launched 
Chandra and XMM, which carry sophisticated grazing incidence mirrors mounted 
aboard sophisticated satellites, include ToO observations of X-ray 
GRB afterglows with subarcminute error boxes. These X-ray telescopes can provide X-ray positions with arcsecond error boxes.  A case in point is the recent 
Chandra observation of GRB000210 on the basis of the sole 
GRB error box provided by the Beppo-SAX wide field camera (Garcia et al. 2000 in 
GCN 544).

Studies of the galaxy 
hosts of GRBs with future 
instruments (e.g. SIRTF, FIRST, NGST, ALMA, etc) 
will open a new cosmic window 
for galaxy 
formation and evolution.

\vskip .05in 
\centerline{\bf Conclusion}
\vskip .05in 

Gamma-ray bursts can be used as beacons to:

\noindent1) study the history of massive star formation in the universe,

\noindent2) to detect the putative population III stars at redshifts $\geq$ 5,  

\noindent3) to detect a population of dust-enshrouded massive galaxies in formation,

\noindent4) to determine the chemical evolution of the universe.

\vskip .05in

Besides large optical telescopes, this approach will require infrared and 
sub-millimeter instruments of unprecedented sensitivity.

\vskip .05in

{\it Acknowledgements:} I.F.M. acknowledges support from Conicet/Argentina, 
and thanks Jacques Paul and Paolo Goldoni for enlightening discussions.

\vskip .2in

\centerline{\bf References}

\vskip .1in

Adelberger, K.L. \& Steidel, C.C. 2000, ApJ, in press (astro-ph 0001126)

Barger, A.J., et al. 1998, Nature, 394, 241

Blain A.W.,  et al. 1999, MNRAS, 302, 632 

Blain, A.W. \& Natarajan, P. 1999, in press (astro-ph/9911468)

Bromm, V., Coppi, P.S. \& Larson, R.B. 1999, ApJ, 527, L5

Carrili, C.L., et al. 2000, ApJ, in press (astro-ph 0002386)

Castro-Tirado A.,  et al. 1999, Science, 283, 2069 

Costa E., et al. 1997, Nature, 387, 783 

Djorgovski S.G., et al. 1999, AAS, 194, 414 

Elbaz D., et al. 1999, A\&A, 351, 37

Fabian, A.C., et al. 1999, MNRAS, 308, L39 

Fishman G.J. 1998, in The Hot Universe, eds. Koyama et al., IAU S188, 159 

Fruchter A.S. 1999, ApJ, 512, L1 

Galama T.J., et al. 1999, Nature, 398, 394  

Greiner J. 1999, GRB web page http://www.aip.de/\~{}jcg 

Hughes, D., et al. 1998, Nature, 394, 241

Kulkarni S.R., et al. 1999, Nature, 398, 389 

Lamb, D.Q. \& Reichart, D.E. 1999, (astro-ph/9909002)

Madau P., et al. 1996, MNRAS, 283, 1388 

MacFayden, A.I. \& Woosley, S.E. 1999, ApJ 524, 262

M\'esz\'aros P. 1999, Nature, 398, 368 

Mirabel I.F. \& Rodr\'\i guez L.F. 1999, ARAA, 37, 409

Mirabel, I.F.,  et al. 1999, V Huntsville Conference on Gamma-Ray Bursts,  
(astro-ph/9912446)

Mushotzky, R.F., Cowie, L.L., Barger, A.J. \& Arnaud, K.A. 2000, Nature, in press

Paczy\'nski B. 1998, ApJ, 494, L45 

Sanders, D.B. 2000, Galaxy Morphology Conf., ed. Block et al., (astro-ph/9910028)

Sanders D.B. \& Mirabel I.F. 1996, ARAA, 34, 749 

Smail I., Ivison R.J. \& Blain A.W. 1997, ApJ, 490, L5 

Steidel C.C., et al. 1999, ApJ, 519, 1 

Taylor G.B., et al. 1998, ApJ, 502, L115 

Woosley, S.E. 1999, V Huntsville Conference on Gamma-Ray Bursts. 
(astro-ph/9912484)

\end